\documentstyle[aps,prl,twocolumn,floats,epsfig]{revtex}

\newcommand{\be}{\begin{equation}}
\newcommand{\ee}{\end{equation}}
\newcommand{\bea}{\begin{eqnarray}}
\newcommand{\eea}{\end{eqnarray}}
\def\bma{\begin{mathletters}}
\def\ema{\end{mathletters}}

\newcommand{\ket}[1]{ | \, #1  \rangle}
\newcommand{\bra}[1]{ \langle #1 \,  |}

\newcommand{\abs}[1]{ | \, #1 \,  |}

\newcommand{\calw}{\mbox{$\cal W$}}
\newcommand{\one}{\mbox{$1 \hspace{-1.0mm}  {\bf l}$}}

\begin{document}
\draft
\wideabs{

\title{Classification of mixed three--qubit states}
\author{A.~Ac\'\i n$^1$, D.~Bru\ss $^2$, 
M.~Lewenstein$^2$, and A.~Sanpera$^2$}
\address{
$^1$ Departament d'Estructura i Constituents de la Mat\`eria, 
Universitat de Barcelona, 08028 Barcelona, Spain \\
$^2$ Institut f\"ur Theoretische Physik, Universit\"at Hannover, 30167
Hannover, Germany}
\date{Received \today}
%\widetext

\maketitle
\begin{abstract}

We introduce a classification of mixed three--qubit states,
in which we define the classes of separable, biseparable, W-- and
GHZ--states. These classes are successively embedded into each other. 
We show that contrary to pure W--type states, the mixed $W$--class
%, contrary to pure W--type states, 
is not of measure zero. 
We construct witness operators that detect the class of a mixed
state. We discuss the conjecture that all entangled states with
positive partial transpose (PPTES) belong to the
$W$--class. Finally, we present a new family of PPTES ``edge" states 
with maximal ranks.
\end{abstract}

}

\narrowtext
The rapidly increasing interest in quantum information processing has
motivated the detailed study of entanglement. 
Whereas entanglement of pure bipartite systems is well understood, 
the classification of mixed states according to the degree 
and character of their entanglement is still a matter of intensive
research (see \cite{primer}).
It was soon realised, that the entanglement of pure tripartite quantum 
states is not a trivial extension of the entanglement of bipartite
systems\cite{ghz,rew2000}. 
Recently, the first results concerning the entanglement of
pure tripartite systems have been achieved\cite{barcel,sudbery,W}.
%has been recently studied in Refs. \cite{barcel,sudbery,W,rew2000}. 
There, the main goal has been to
generalize the concept of the Schmidt decomposition to three-party systems 
\cite{barcel,sudbery}, 
and to distinguish classes of locally inequivalent 
states\cite{W}. The knowledge of mixed 
tripartite entanglement is much less advanced
(see, however, \cite{upb,mix,jenshans}).

In this Letter we introduce a classification of the whole
space of mixed three--qubit states into different entanglement
classes. We provide a method to determine to which class a given state 
belongs (tripartite witnesses). We also discuss the characterization of 
 entangled states that are positive under partial transposition 
(PPTES). Finally, we introduce a new family 
of PPTES for mixed tripartite qubits. 
 
Our proposal to classify mixed tripartite--qubit states 
is done by specifying compact convex subsets of the space of all states,
which are  embedded into each other.
This idea vaguely resembles the classification of bipartite
systems by their Schmidt number \cite{jenshans,terhor,us}. However,
as shown later our classification does {\em not} 
follow the Schmidt number\cite{jenshans}.
%as introduced in 
%\cite{jenshans}. 
Also in this respect, entanglement 
of tripartite systems differs genuinely from the one of
bipartite quantum systems.

Before presenting our results concerning mixed states, we briefly review
some of the recent results on pure three--qubit states. 
Any three--qubit vector (pure state)  can be written as 
\bea
\ket{\psi_{GHZ}} &=&
\lambda_0\ket{000}+\lambda_1e^{i\theta}\ket{100}+\lambda_2\ket{101}
\nonumber\\&+&\lambda_3\ket{110}+\lambda_4\ket{111}\ ,
          \label{GHZ}
\eea
where $\lambda_i\geq 0$, 
%$0\leq\theta\leq\pi$, 
$\sum_i\lambda_i^2=1$,
$\theta\in [0,\pi]$
,
and $\{|0\rangle,
|1\rangle\}$ denotes an orthonormal basis in Alice's, Bob's and
Charlie's space, respectively\cite{barcel}. 
Apart from separable and biseparable pure states, 
there exist also two different types of locally inequivalent 
entangled vectors; the so-called 
GHZ--type \cite{ghz} and W--type\cite{W}.
Vectors belonging to GHZ-- and W--types cannot be transformed  into 
each other 
by local operations and classical communication (LOCC). 
Generically, a vector described by Eq.(\ref{GHZ}) is of the GHZ--type, 
while W--vectors can be written as
 \be
 \ket{\psi_{W}} = \lambda_0\ket{000}+\lambda_1\ket{100}
          +\lambda_2\ket{101} +\lambda_3\ket{110} .
 \label{W}
 \ee 
W--vectors  form a set of measure zero among 
all pure states\cite{W}. Also, given a W--vector  
one can always find a GHZ--vector as close  
to it as desired by adding an
infinitesimal $\lambda_4$--term to the RHS of Eq.(\ref{W})
\cite{note1}. Furthermore, the so-called tangle, $\tau$, 
introduced in \cite{wootters}, can be used to detect  the type, since
$\tau(\ket{\psi_{W}})=0$ \cite{W}.

\begin{figure}[ht]
\vspace*{-2.2cm}
\hspace*{2cm}
\begin{picture}(150,150)
\put(5,5){\epsfxsize=140pt\epsffile[23 146 546 590 ]{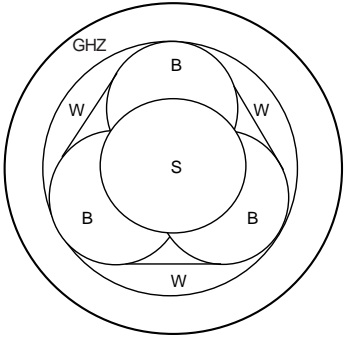}}
\end{picture}
\vspace*{1.4cm}
\caption[]{\small  Schematic structure of the set of all three--qubit states.
        $S$: separable class; $B$: biseparable class (convex hull
of biseparable states with respect to any partition); 
         $W$--class and
        $GHZ$--class.}
\label{Fig.1}
\end{figure}

Mixed states of three--qubit systems can be classified 
generalizing the classification of pure states. 
To this aim we
define (see Fig.\ref{Fig.1}):\\
\indent$\bullet$ the class $S$ of separable states, i.e. those that can be expressed
as a convex sum of projectors onto product vectors;\\
\indent $\bullet$ the class $B$ of biseparable states, i.e.  
those that can be
expressed as a convex sum of projectors onto product and
bipartite entangled vectors (A--BC, B--AC and C--AB);\\
\indent $\bullet$ the class $W$ of W--states, i.e. those that can be
expressed as a convex sum of projectors onto product, biseparable 
 and W--type vectors;\\
\indent $\bullet$ the class $GHZ$ of GHZ--states, i.e. the set of all
physical states.\\
\noindent All these sets are convex and compact, and satisfy $S\subset B\subset 
W\subset GHZ$. States in $S$ are not entangled.
No genuine three--party entanglement is needed  to prepare entangled
states in the subset $B\setminus S$. The formation of
entangled states in $W\setminus B$  requires  
W--type vectors with three--party entanglement,  but zero tangle, which
is an entanglement monotone decreasing under  LOCC \cite{W}.  Finally, the class
$GHZ$ contains all types of entanglement, 
%the types of entanglement are necessary in order 
%to form the states from $GHZ$,
and in particular,   GHZ--type vectors are needed to prepare 
states  from  $GHZ\setminus W$. 
The introduced classes are invariant under local unitary or
invertible non-unitary operations,  while local POVM's
\cite{note1} can only transform 
states from a ``higher" to a ``lower" class.
%%%%%%%%%%%%%%%%%%%%%%%%%%%%%%%%%%%%%%%%%%%%%%%%%%%%%%%%%%%%%%%%%%%%
%%%%%%%%%%%%%%%%%%%%%%%%%%%%%%%%%%%%%%%%%%%%%%%%%%%%%%%%%%%%%%%%%%%%%%%%

Notice that since GHZ--vectors
can be expressed as the sum of only two product vectors, i.e. 
$\ket{GHZ}=(\ket{000}+\ket{111})/\sqrt{2}$, whereas the minimum
number of product terms forming a W--vector is  three
\cite{barcel,W}, 
as in the state 
$\ket{W}=(\ket{100}+\ket{010}+\ket{001})/\sqrt{3}$, 
our scheme may seem somehow counterintuitive.
Indeed, for bipartite systems, states with lower Schmidt number, i.e. lower
number of product terms in the Schmidt decomposition, are embedded into 
the set of states with higher Schmidt number \cite{terhor}.
One is tempted to extend this classification to tripartite systems
as  $S\subset B\subset GHZ  \subset W$, where now $W$ is the set of all states.
However, such generalization is  evidently wrong, because the 
the  set of GHZ--states in such classification cannot be closed \cite{note1}.

%As we demonstrate, this cannot be generalized to tripartite
%states. 
%Indeed, a naive generalization would mean to classify the states 
%as $S\subset B\subset GHZ 
%\subset W$, where now $W$ is the set of all states. 
%Such generalization is  evidently wrong, because the  set of GHZ--states in the
%above classification cannot be closed (\cite{note1}).

Having established the structure of the set of mixed three--qubit states,
we show how to determine to which class a given 
state $\rho$ belongs. To this aim, we use the approach developed 
previously in the construction and optimisation of witness operators 
\cite{us,terhal,witness}.
% method of subtracting projectors onto  vectors from
%a given subset \cite{BSA},  and witness operators.
%Notice that the tangle,
%$\tau$, can be defined for mixed states as
%\be
%\label{tangle}
%\tau(\rho)=\min\sum_i p_i\tau(\ket{\psi_i}) ,
%\ee
%where the minimum is with respect to all the decompositions  
%$\rho=\sum_ip_i\ket{\psi_i}\bra{\psi_i}$. This quantity would be 
%very useful for detecting GHZ--states, since they have 
%$\tau\neq 0$, but a closed expression for (\ref{tangle}) is
%not known \cite{wong}.

We denote the range of $\rho$ 
by $R(\rho)$, its rank by $r(\rho)$, its kernel by $K(\rho)$, and the
dimension of $K(\rho)$ by $k(\rho)$.
%and its dimension as  $k(\rho)$.
Following the approach of the best separable
approximation (BSA)\cite{BSA},
one can decompose any state $\rho$
as a convex combination  of a $W$--class state and a remainder $\delta$,
\be
\rho = \lambda_W \rho_W +(1-\lambda_W)\delta\ ,
\label{bsa}
\ee
where 
$0\leq \lambda_W \leq 1$, and $R(\delta)$ does not contain 
any W--vector. Maximization of $\lambda_W$ leads
to the best W--approximation of $\rho$. Notice that  only for $
\rho$ belonging to the $GHZ\setminus W$--class, this decomposition is  
non-trivial, i.e. $\lambda_W\neq 1$. Also, $r(\delta) = 1$,
since any subspace spanned by two linearly independent 
GHZ--vectors contains at least one pure state with zero tangle. 
In fact, given $\ket{\psi_1}$ and $\ket{\psi_2}$ 
with $\tau(\ket{\psi_1})$ 
and $\tau(\ket{\psi_2})$ 
not equal zero, it is always possible to find some $\tilde\alpha,\tilde\beta$ 
such that
$\ket{\psi(\tilde\alpha,\tilde\beta)}=\tilde\alpha\ket{\psi_1}+
\tilde\beta\ket{\psi_2}$ 
is normalized, and its tangle is zero.
Therefore, any W--approximation must have the form:
\be
\rho = \lambda_W \rho_W +(1-\lambda_W)\ket{\psi_{GHZ}}
    \bra{\psi_{GHZ}}\ .
\ee
%where $\ket{\psi_{GHZ}}$ denotes a pure state from $GHZ\setminus W$.
%In a similar way one can write  $\rho$ in a
Similarly, one can express $\rho$ in the best
biseparable approximation as:
\be
\rho = \lambda_B \rho_B +(1-\lambda_B)\delta\ ,
\ee
where now $R(\delta)$ must not contain any biseparable states, i.e.
$r(\delta)<4$, 
%This inequality follows from the fact that
since any $N$--dimensional subspace of the  $2\times N$ space 
contains at least one product vector \cite{2byn}. 

We use the  above decompositions to construct operators
that detect the desired  subset (see \cite{witness}).  
%will call tripartite witnesses, i
In analogy to entanglement witnesses 
%\cite{terhal}, 
and Schmidt witnesses 
%\cite{us} 
we term these operators
tripartite witnesses. 
The existence  of  
witness operators is a consequence of the Hahn-Banach theorem, which states that 
a point outside a convex compact set is separated from that set by a 
hyper-plane. The equation Tr$(\calw \rho)=0$
describes such a hyper-plane, and one calls
$\calw$ a witness operator. For example,
in our setting, a W--witness is an operator
$\calw_W$ such that Tr$( \calw_W\rho_B)\geq 0$ holds 
$\forall \rho_B\in B$, but for which
there exists a $\rho_W\in W \setminus B$ such that Tr$( \calw_W\rho_W)<0$.

Any GHZ--witness (W--witness) has the canonical form $\calw=Q-\epsilon \one$, 
where $Q$ is a positive operator which has no W--type (B--type) vectors in 
its kernel; thus $k(Q)=1$ ($k(Q)<4$)\cite{us,witness}. 
An example of a GHZ--witness is
\be
\calw_{GHZ} = \frac{3}{4}\one-P_{GHZ}\ ,
\ee
where $P_{GHZ}$ is the projector onto  
$\ket{GHZ}$.
%=(\ket{000}+\ket{111})/\sqrt{2}$. 
The value 3/4 corresponds
to the maximal squared overlap between  $|GHZ\rangle$ and a
W--vector.
% and a $|GHZ\rangle$.
This construction guarantees that Tr$(\calw_{GHZ}\rho_W)\geq 0$ 
for any W--state, and since  Tr$(\calw_{GHZ}P_{GHZ})< 0$, there is 
a GHZ $\setminus$ W--state which is detected by $\calw_{GHZ}$. 
The maximal overlap 
%between a W--vector and  $\ket{GHZ}$ 
is obtained as follows: due to the symmetry of $\ket{GHZ}$
we  only need to consider W--vectors that are symmetric under the
exchange of any of the three qubits\cite{note2}. Therefore, we 
 have to consider
all local trilateral rotations of
$\ket{\psi_{W}}=\kappa_0\ket{000}+\kappa_1(\ket{100}+\ket{010}+\ket{001})$,
where $\kappa_0, \kappa_1$ are real and $\kappa_0^2+ 3\kappa_1^2=1$.
Due to the symmetry, 
such rotations can be   parametrised for all parties as
 $\ket{0}\rightarrow \alpha\ket{0}+\beta\ket{1}$,
$\ket{1}\rightarrow \beta^*\ket{0}-\alpha^*\ket{1}$, with 
$\abs{\alpha}^2+\abs{\beta}^2=1$. 
Thus, the overlap $\bra{GHZ}\psi_W\rangle$ is a function of 
six parameters with two constraints, and can be maximized using
%with the method of 
Lagrange multipliers. An optimal choice of parameters is  $\kappa_0=0$,
$\kappa_1=1/\sqrt{3}$, and $\beta=-\alpha=1/\sqrt{2}$.
This leads to $\abs{\bra{GHZ}\psi_W\rangle}^2_{max}=3/4$. 

Analogously, we can construct a  
W--witness as
\be
\calw_{W_1} = \frac{2}{3}\one-P_{W}\ ,
\label{wwitness}
\ee
where $P_{W}$ is now the projector onto a vector $\ket{W}$, and 2/3 
corresponds to the maximal squared overlap  between 
$\ket{W}$ and a B--vector. 
%The overlap if found in the same
%way as before. 
%Again we found the maximal overlap using a same method as before.  
%consider all local rotations of the normalized B--vector 
%$\ket{\psi_{B}}=\ket{0}(\alpha \ket{10}+\beta \ket{01})$,
%where, due to symmetry of the W state for the second and third bit, the local
%rotations for the second and third bit of $\ket{\psi_{B}}$ are identical.
%%The overlap $\bra{\psi_{W}}\psi_B\rangle$ therefore depends on 3 complex
%parameters (after consideration of constraints due to normalization),
%The optimal parameters are 
%$\beta=\alpha=1/\sqrt{2}$, and no
%local rotation. 
Another example of a W--witness is
\be
\calw_{W_2} = \frac{1}{2}\one-P_{GHZ},
\label{wwitness2}
\ee
where now 1/2 is the maximal squared overlap between $\ket{GHZ}$
and a B--type vector \cite{dur}.
% with a  $\ket{GHZ}$.
% this is shown following a similar method as above.
The W--vector that has maximal overlap
with  $\ket{GHZ}$ is detected by $\calw_{W_2}$.

The tripartite witness  $\calw_{W_2}$
%given  (\ref{wwitness2})
allows to prove that the class of mixed 
W $\setminus$ B--states is not of measure zero:
%although the set of pure W--type states 
%is of measure zero among the pure states 
%\cite{W}.
% does not  mean that the same holds for the
%set of mixed states of the $W$--class.
%To this aim, 
consider the 
%following 
family of states
in ${\mathcal C}^2\otimes {\mathcal C}^2\otimes {\mathcal C}^2$
given by  the convex sum of the identity and a projector onto a 
W--state,
\be
\label{measst}
\rho = \frac{1-p}{8}\one + p P_W\ .
\ee
Obviously, the states (\ref{measst}) belong at most to $W$.
%, and in fact their 
%tangle (\ref{tangle}) is equal to zero.
The range for the parameter $p$, in which  $\calw_{W_2}$
%$\calw = 1/2-P_{GHZ}$ from (\ref{wwitness2}) 
detects $\rho$, i.e.
Tr$(\calw_{W_2} \rho)< 0$,  is found to be $3/5 < p \leq 1$,
and is bigger than the one
found by using  $\calw_{W_1}$. 
%With respect to $\rho$, 
%$\calw_{W_2}$ is ``finer'' than
%$\calw_{W_1}$.
Taking any $p$ which has a finite distance to the
border of this interval, i.e. $p-3/5>\Delta$ and
$1-p>\Delta$, it is  always possible
to find a finite region around $\rho$ which still belongs
to the $W \setminus B$--class. This can be seen by considering
\be
\tilde \rho = (1-\epsilon)\left[\frac{1-p}{8}\one + p P_W\right]+
\epsilon \sigma\ ,
\ee
where $\sigma$ is an {\em  arbitrary} density matrix, which covers all
directions of  possible deviations from $\rho$ in the operator space. 
In the worst case
$\sigma$ is orthogonal to $P_{GHZ}$, so that Tr$(P_{GHZ}\sigma)=0$, and 
therefore
Tr$(\calw_{W_2} \tilde \rho)=(1-\epsilon)\text{Tr}(\calw_{W_2}  \rho)+
\epsilon/2$.
As long as 
the relation
$\epsilon <(5p-3)/(5p+1)$ holds, 
%i.e. for small $\Delta$, $\epsilon\le 5
%\Delta/6$, 
the corresponding  state $\tilde\rho$ is still detected by
$\calw_{W_2}$. 
Moreover, one can also find a 
finite 
$\epsilon'$ such that if $\epsilon<\epsilon'$, then %$\tau(\tilde \rho)=0$.
$\tilde \rho$ is in the $W$--class. The bound $\epsilon'$ is obtained, 
for instance, by demanding that 
$(1-\epsilon')(1-p)\one/8+\epsilon'\sigma$ is biseparable. 
The intersection of the two
intervals gives a finite range for 
$\epsilon$ where the state $\tilde \rho$ is in the $W \setminus B$--class.
This proves that the set of mixed W $\setminus$ B--states contains a ball, i.e. 
is not of measure zero. 

We discuss now some possible  consequences of our results for  
PPTES of three qubits, for which the partial transposes  $\rho^{T_A}$, 
$\rho^{T_B}$ and $\rho^{T_C}$ are positive. 
Any of these states can be decomposed as:
\be
\rho = \lambda_S \rho_S +(1-\lambda_S)\delta\ ,
\label{pptbsa}
\ee
where $\rho_S$ is a separable state and $\delta$ is an edge state\cite{sini}.
We conjecture that {\it PPTES
cannot belong to the GHZ $\setminus$ W--class},  
i.e. they are at most in the $W$--class.
This conjecture  is rigorous for states that have 
edge states with low ranks in the above decomposition.
It was shown in \cite{2byn} that for bipartite systems
 in
${\mathcal C}^2\otimes {\mathcal C}^N$, 
the rank of PPTES must be larger than $N$,
and  if $r(\rho)\le N$ and $\rho^{T_A}\geq 0$, then
the state $\rho$ is separable.
Thus, any PPTES of three--qubits with
$r(\rho)\leq 4$ is biseparable with respect to any
partition; an example of such states are  the UPB--states from Ref. \cite{upb}. 

For the case of  higher ranks we can only give some support for our 
conjecture. We proceed as in \cite{us}, and observe first that it suffices to 
prove the conjecture for the edge states. 
For these states, the sum of ranks
satisfies $r(\delta)+r(\delta^{T_A})+r(\delta^{T_B})+r(\delta^{T_C})\leq 28 $\cite{sini}.
Any PPT entangled state can only be  detected by a non-decomposable 
entanglement witness, which in the case of tripartite systems has the canonical form 
$\calw_{nd}=\calw_d-\epsilon\one$ where $\calw_{d}=P+
\sum Q^{T_X}_X$ is a decomposable operator with $P,Q_X \geq 0$,   
$R(P)=K(\delta)$, $R(Q_X)=K(\delta^{T_X})$ for some edge state $\delta$, 
and $X=A,B,C$
\cite{sini}.  
%As in Ref. \cite{us} 
We restrict ourselves to edge states with the maximal sum of
ranks, i.e. states 
$\delta$ with $(r(\delta),r(\delta^{T_A}),r(\delta^{T_B}),r(\delta^{T_C}))
=(8,8,7,5),(8,8,6,6),(8,7,7,6),(7,7,7,7)$ and permutations. Indeed, if the
conjecture is true for these states, it will be true for all  edge
states, and thus for all PPTES, 
since the edge states with maximal sum of ranks are
dense in the set of all edge states \cite{us}. 
We conjecture that for the case of edge states 
with maximal sum of ranks it is always possible 
to find a pure W--type vector, $\ket{\phi_W}$, such that for any 
non-decomposable witness $\calw_{nd}$ of $\delta$, 
$\bra{\phi_W}\calw_d\ket{\phi_W}\leq 0$, so that 
$\bra{\phi_W}\calw_{nd}\ket{\phi_W}< 0$.  That means $\calw_{nd}$ 
{\it cannot} be a GHZ--witness, so
%In turn, this implies that 
the edge state $\delta$ belongs to the 
$W$--class. 
If this holds for any $\delta$ it implies that all 
PPTES belong to the $W$--class.\\  
Any W--vector can be obtained by local invertible 
operations applied to
$\ket{W}$
%=(\ket{100}+\ket{010}+\ket{001})/\sqrt{3}$, 
i.e. can be written as 
$\ket{\phi_W}=\alpha_A\ket{e_2,f_1,g_1}+\alpha_B\ket{e_1,f_2,g_1}
+\alpha_C\ket{e_1,f_1,g_2}$. We denote
$\ket{\Phi_A}=\ket{e_2^*,f_1,g_1}$, $|\Psi_A\rangle=\alpha_B\ket{e^*_1,f_2,g_1}
+\alpha_C\ket{e^*_1,f_1,g_2}$, $\ket{\Phi_B}=\ket{e_1,f^*_2,g_1}$,
$|\Psi_B\rangle=\alpha_A\ket{e_2,f^*_1,g_1} +\alpha_C\ket{e_1,f^*_1,g_2}$, 
$\ket{\Phi_C}=\ket{e_1,f_1,g^*_2}$,
$|\Psi_C\rangle=
\alpha_A\ket{e_2,f_1,g^*_1}+\alpha_B\ket{e_1,f_2,g^*_1}$. In order to fulfill
the condition  $\bra{\phi_W}\calw_d\ket{\phi_W}\leq 0$  we demand that
$Q_X|\Phi_X\rangle=0$; 
%Q_B\ket{\Phi_B}= Q_C\ket{\Phi_C}=0$;
$P\ket{\phi_W}=0$, and  $Q_X|\Psi_X\rangle=0$ for $X=A,B,C$. 
The latter 4 conditions form 4 linear homogeneous equations 
for the $\alpha_X$'s, whose solutions exist if 
two $3\times 3$ determinants vanish. 
Together with the first 3 conditions this gives at most 5 equations  in
the case $r(\delta)<8$, and 6 equations in the worst case $r(\delta)=8$,  
for the 6 complex parameters characterizing 
$\ket{e_i}, \ket{f_i}$, and $\ket{g_i}$, with $i=1,2$.
For $r(\delta)<8$ ($r(\delta)=8$) one expects here a
one complex parameter (finite, but
large) family of solutions. At the same time
$\bra{\phi_W}\calw_d\ket{\phi_W}= 2\,{\rm Re}\sum_X 
\alpha_X \bra{\Psi_X^{*_X}}Q_X^{T_X}\ket{\Phi_X^{*_X}}$, 
(where $\ket{\Phi^{*_X}}$ denotes partial complex
conjugation with respect to $X$)
i.e. is a
hermitian form of $\alpha_X$'s, whose diagonal
elements vanish, 
since $\ket{\Psi_X}$ does not depend on $\alpha_X$.  Employing the freedom 
of choosing the solutions from the family, one expects to find at least one with 
$\bra{\phi_W}\calw_d\ket{\phi_W}\le 0$. 
In this way we obtain the W--vector we were
looking for.  For the cases $(6,8,8,6)$ and 
$(5,8,8,7)$, a similar argument indeed shows that there should exist 
a biseparable state, $\ket{\psi_B}$, such that 
$\bra{\psi_B}\calw_{nd}\ket{\psi_B}<0$. Note that the above
method of searching $\ket{\psi_W}$ ($\ket{\psi_B}$) for a given $\delta$,  if
successful, provides a sufficient condition  for $\delta$ to belong to the 
$W$--class
($B$--class).

Finally,  we present an example for a PPTES entangled edge state 
with ranks (7,7,7,7).
We introduce
\be
\rho = \frac{1}{n}\left(\begin{array}{cccccccc} 
                  1&0&0&0&0&0&0&1 \\
                  0&a&0&0&0&0&0&0 \\
                  0&0&b&0&0&0&0&0 \\
                  0&0&0&c&0&0&0&0 \\
                  0&0&0&0&\frac{1}{c}&0&0&0 \\
                  0&0&0&0&0&\frac{1}{b}&0&0 \\
                   0&0&0&0&0&0&\frac{1}{a}&0 \\ 
                   1&0&0&0&0&0&0&1 
                \end{array} \right)\ \  
\ee
with $a,b,c >0$ and $n=2+a+1/a+b+1/b+c+1/c$.
The basis  is $\{000,001,010,011,100,101,110,111 \}$. This 
density matrix
has a positive partial transpose with respect to each subsystem. One sees
immediately  that $r(\rho)=r(\rho^{T_A})=r(\rho^{T_B})=
r(\rho^{T_{AB}})=7$. In order to  check that  $\rho$ is a 
PPT entangled edge state, one has to prove that it is impossible
to find a product vector $\ket{\phi}\in R(\rho)$, such that at the same time
$\ket{\phi^{*_X}}\in R(\rho^{T_X})$ for $X=A,B,C$.
%$\ket{\phi^{*_B}}\in R(\rho^{T_B})$, and 
%$\ket{\phi^{*_C}}\in R(\rho^{T_C})$.
This, indeed, is not possible, as one readily concludes 
by looking at the kernels directly: one cannot find a product
vector $\ket{\phi}$ that is orthogonal to $\ket{000}-\ket{111}$, 
whereas at the same time 
$\ket{\phi^{*_A}}\perp \ket{011}-c\ket{100}$,
$\ket{\phi^{*_B}}\perp \ket{010}-b\ket{101}$,
and $\ket{\phi^{*_C}}\perp \ket{001}-a\ket{110}$, unless the condition 
$ab=c$ is fulfilled. Thus, for generic $a,b,c$ we have found a family of 
bound PPT entangled edge states of three qubits with maximal sum of ranks. 
By direct inspection we observe 
that $\rho$
fulfills our conjecture, and is  biseparable with respect to any 
partition. 
It can be written e.g. as a sum of separable
projectors and a B--state acting in the $2\times 2$ subspace spanned by 
Alice's space and the vectors $\ket{00}$ and $\ket{11}$ in
Bob's--Charlie's space.
% (it is indeed biseparable for all possible partitions).

To summarize, we show that the set of  density matrices for three qubits
has an ``onion" structure (see Fig.1) and contains convex compact subsets
of states belonging to  the  separable $S$, biseparable $B$, $W$-- and
$GHZ$--class, respectively. 
We provide  the  canonical way of
constructing witness operators for the $GHZ$-- and $W$--class,
and give the first examples of such witnesses.
%present the   construction of
%necessary conditions for a given state to belong to the
%W--, or B--class, and give  
%first examples. 
The  study of %generalized Werner states \cite{werner} 
the family of tripartite states given in Eq. (\ref{measst}
)
allows us to  prove that the $W$--class is not of measure zero. 
We conjecture and give  some evidence that all PPTES 
of three--qubit systems do not require GHZ--type
pure states for their formation. We  formulate 
a sufficient condition  which allows to check
constructively if a state  belongs to the $W$--class
($B$--class). Finally, we present a family of  PPT entangled edge states
of three qubits with maximal sum of ranks.

This work has been supported  by  DFG 
(SFB 407 and Schwerpunkt  ``Quanteninformationsverarbeitung"), 
the ESF-Programme PESC,  and the EU IST-Programme
EQUIP. AA thanks the University of Hannover for 
hospitality, E. Jan\'e for useful comments and the 
Spanish MEC (AP-98)  for financial support.

%here are the figures:
%\newpage

%\end{multicols}

\begin{thebibliography}{99}
\bibitem{note1} This implies that a GHZ--vector can be {\em approximately} 
transformed by LOCC into a W--vector with some probability, although the 
contrary is, in general, not possible.
\bibitem{note2} One can always symmetrize an asymmetric state without 
diminishing the overlap.
\end{thebibliography}
\end{document}